\documentclass[fleqn]{elsart}
\usepackage{ifpdf,graphicx,lineno,epsfig,bm,amsfonts,amssymb,amsmath}
\begin{document}

\begin{frontmatter}

\title{Centrifugal deformations of the \\ gravitational kink}

\author[it]{Paolo~Maraner},
\author[uk]{Jiannis~K.~Pachos}

\address[it]{School of Economics and Management, Free University of Bozen-Bolzano,\\ via Sernesi~1, 39100~Bolzano, Italy}
\address[uk]{School of Physics and Astronomy, University of Leeds, Leeds LS2 9JT, UK}

\begin{abstract}
The Kaluza-Klein reduction of 4d conformally flat spacetimes is
reconsidered. The corresponding 3d equations are shown to be equivalent to
2d gravitational kink equations augmented by a centrifugal term.
For space-like gauge fields and
non-trivial values of the centrifugal term the gravitational kink solutions
describe a spacetime that is divided in two disconnected regions.
\end{abstract}


\end{frontmatter}

\section{Introduction}
 \label{Intro}
Reducing the gravitational Chern-Simons term from three to two
dimensions produces an interesting as well as surprising outcome
\cite{Guralnik&Iorio&Jackiw&Pi03}. On the one hand the resulting
equations of motion require the vanishing of the 3d Cotton tensor and,
hence, impose conformal flatness of the original 3d spacetime. On the other
hand, they take the form of general covariant kink equations that, besides
symmetry breaking solutions, support an associated kink. Let us parameterize
the 3d line element as $ds_{(3)}^2=\textrm{g}_{ij}dx^i dx^j +(\textrm{A}_i
dx^i+ dx^2)^2$, where Latin indices range over $0,1$ and all quantities are
independent of the third coordinate $x^2$. Then, the equations obtained by
Guralnik, Iorio, Jackiw and Pi ((4.47,48,49) in
Ref.~\cite{Guralnik&Iorio&Jackiw&Pi03}) read
\begin{subequations}
\begin{eqnarray}
&&\textrm{R}=k+3f^2,\label{gk2dEq(a)}\\
&&\textrm{D}^2f+kf-f^3=0,\label{gk2dEq(b)}\\
&&\textrm{D}_i\textrm{D}_jf-\mbox{$\frac{1}{2}$}\textrm{g}_{ij}\textrm{D}^2f=0.\label{gk2dEq(c)}
\end{eqnarray}\label{gk2dEq}\end{subequations}
Here $\textrm{g}_{ij}$ is the 2d spacetime metric, $\textrm{D}_i$
is the associated covariant derivative,
$\textrm{D}^2=\textrm{D}_i\textrm{D}^i$, $\textrm{R}$ is the
relative scalar curvature and $k$ is an arbitrary constant. The
Kaluza-Klein field strength $\textrm{F}_{ij}=\partial_i
\textrm{A}_j-\partial_j \textrm{A}_i$ has been expressed in terms
of its dual scalar\footnote{Some signs differ from
Ref.~\cite{Guralnik&Iorio&Jackiw&Pi03}, and in what follow from
Ref.~\cite{Grumiller&Jackiw06}, due to different conventions. Here
${\textrm{R}_{ijk}}^l=\partial_i\Gamma_{jk}^l-...$,
$\textrm{R}_{ij}={\textrm{R}_{kij}}^k$, $\textrm{R}={\textrm{R}_i}^i$
and analogously in other dimensions. The single time-like direction carries
negative signature, while space-like directions carry positive signature.}
\begin{equation}
\textrm{F}_{ij}\equiv\sqrt{-\textrm{g}}\varepsilon_{ij}\,f
\end{equation}
Equation (\ref{gk2dEq(b)}) is the covariant generalization of the
flat space kink equation with biquadratic potential
\begin{equation}
V(f)=\mbox{$\frac{1}{4}$}(k-f^2)^2,\label{VKink}
\end{equation}
supporting, for positive values of $k$, the well known kink
profile. Equations (\ref{gk2dEq(a)}) and (\ref{gk2dEq(c)}) appear
instead, as subsidiary constraints necessary to uniquely determine
the geometry $\textrm{g}_{ij}$ together with the field $f$. As shown in
Ref.~\cite{Guralnik&Iorio&Jackiw&Pi03}, general covariant kinks
bear a close relation to flat space kinks governed by identical
potentials. Indeed, equations (\ref{gk2dEq}) support the solution
\begin{subequations}
\begin{eqnarray}
&&\textrm{g}_{ij}dx^idx^j=-\mbox{$\frac{1}{4}$}k^2\,\mbox{sech}^4
\mbox{$\left(\frac{\sqrt{k}}{2}x^1\right)$}(dx^0)^2+(dx^1)^2,\label{gk2dMetric}\\
&&\textrm{A}_i dx^i=\mbox{$\pm\mbox{$\frac{1}{2}$}
k\,\mbox{sech}^2\left(\frac{\sqrt{k}}{2}x^1\right)dx^0$},\label{gk2dVectorPotential}
\end{eqnarray}
with the corresponding kink profile
\begin{equation}
\mbox{$f=\pm\sqrt{k}\tanh\left(\frac{\sqrt{k}}{2}x^1\right).$}
\label{gk2dKinkProfile}
\end{equation}\label{gk2dSolutions}\end{subequations}
It is rather natural to wonder wether this situation generalizes to higher
dimensions. Indeed, the Kaluza-Klein reduction of conformally flat spaces
was recently considered by Grumiller and Jackiw for arbitrary
dimensions~\cite{Grumiller&Jackiw06}. The resulting, rather daunting,
equations dramatically simplify for the reduction from 4d to 3d. Grumiller
and Jackiw restricted to that case and constructed special solutions based
on a further Ansatz on the 3d metric. Here, we reconsider these equations in
their generality. For space-like gauge fields,
we show that the 3d general covariant equations describing
the Kaluza-Klein reduction of 4d conformally flat spaces correspond to the
2d general covariant kink equations (\ref{gk2dEq}) supplemented by a
centrifugal term. Hence, besides the 3d extension of the gravitational kink
(\ref{gk2dSolutions}), 4d conformally flat Kaluza-Klein spaces also support
their `centrifugal deformations'.

Our discussion proceeds as follows. In \S\ref{GJ} we briefly review 3d
Grumiller-Jackiw equations and supplement them with some extra
considerations. In \S\ref{Dc} we take advantage of general covariance to
adapt coordinates and reduce the equations to 2d centrifugal kink equations.
The centrifugal deformations of the gravitational kink are eventually
constructed in \S\ref{gk3d}.

\section{Grumiller-Jackiw equations in 3d}
 \label{GJ}
Let us parameterize the 4d line element as $ds_{(4)}^2=g_{\mu\nu}dx^\mu
dx^\nu +(A_\mu dx^\mu+dx^3)^2$, where Greek indices range over $0,1,2$ and
all quantities are independent of the last coordinate $x^3$. Then, the 3d
Grumiller-Jackiw equations\footnote{In 3d, Grumiller-Jackiw equations
are the Minkowskian analogue of the Euclidean Einstein-Weyl equations, widely
studied and completely solved in mathematics (see \cite{Jackiw07},
\cite{Grumiller&Jackiw08} and references therein).} ((22a,b) in
Ref.~\cite{Grumiller&Jackiw06}) read
\begin{subequations}
\begin{eqnarray}
&&R_{\mu\nu}-\mbox{$\frac{1}{3}$}Rg_{\mu\nu}=
f_\mu f_\nu-\mbox{$\frac{1}{3}$}f_\kappa f^\kappa g_{\mu\nu},\label{cf3dEq(a)}\\
&&D_\mu f_\nu+D_\nu f_\mu=0.\label{cf3dEq(b)}
\end{eqnarray}\label{cf3dEq}\end{subequations}
Here, $g_{\mu\nu}$ is the 3d spacetime metric, $D_\mu$ is the associated
covariant derivative and $R_{\mu\nu}$ and $R$ are the corresponding Ricci
and scalar curvatures. The Kaluza-Klein field strength
$F_{\mu\nu}=\partial_\mu A_\nu-\partial_\nu A_\mu$ has been reexpressed in
terms of its dual vector
\begin{equation}
F_{\mu\nu}\equiv \sqrt{-g}\varepsilon_{\mu\nu\kappa}\,f^\kappa.
\end{equation}
Grumiller and Jackiw also derived two important consequences of equations
(\ref{cf3dEq}a) and (\ref{cf3dEq}b). First, the scalar curvature $R$ can be
expressed in terms of $f_\mu f^\mu$ and an arbitrary constant $k$ as ((23) in
Ref.~\cite{Grumiller&Jackiw06})
\begin{equation}
R=3k-5f_\mu f^\mu. \label{CurvatureConstraint}
\end{equation}
Second, the vector $F^\mu= \frac{1}{\sqrt{-g}}
\varepsilon^{\mu\nu\kappa}D_\nu f_\kappa$, when not identically
vanishing, is a second Killing vector of the geometry  ((26c) in
Ref.~\cite{Grumiller&Jackiw06})
\begin{equation}
D_\mu F_\nu + D_\nu F_\mu=0. \label{2ndKill}
\end{equation}
Before proceeding it is useful to rewrite equation (\ref{2ndKill}) in a
slightly different form and to supplement equations (\ref{cf3dEq}a) and
(\ref{cf3dEq}b) with a further integrability condition.\\ After rising both
indices of (\ref{2ndKill}) we reexpress $F^\mu$ in terms of $f^\mu$
obtaining
\[
\varepsilon^{\lambda\rho\sigma}D^\kappa D_\rho f_\sigma+
\varepsilon^{\kappa\rho\sigma}D^\lambda D_\rho f_\sigma=0.
\]
Contracting with $\varepsilon_{\lambda\mu\nu}$, reexpressing
$\varepsilon$-symbols products in terms of Kronecker deltas and taking into
account (\ref{cf3dEq(b)}) we immediately have
\begin{equation}
D_\kappa D_\mu f_\nu- \mbox{$\frac{1}{2}$} g_{\kappa\mu}\,D^2
f_\nu+ \mbox{$\frac{1}{2}$}g_{\kappa\nu}\,D^2 f_\mu=0,
\label{cf3dTE}
\end{equation}
closely resembling the traceless equation (\ref{gk2dEq(c)}).\\ To
derive the integrability condition consider the covariant
derivative of (\ref{cf3dEq(b)})
\[D_\kappa D_\mu f_\nu+D_\kappa D_\nu f_\mu=0.\]
Antisymmetrizing in $\kappa$ and $\mu$, contracting $\mu$ with $\nu$ and
taking into account the vanishing of the covariant divergence of the gauge
field, $D_\mu f^\mu=0$, we have
 \[\left[D_\kappa,D_\mu\right]f^\mu-D_\mu D^\mu f_\kappa=0.\]
Expressing the commutator of covariant derivatives in terms of the
Ricci tensor and inserting (\ref{cf3dEq(a)}) and
(\ref{CurvatureConstraint}), we eventually obtain the
integrability condition
\begin{equation}
D^2f^\mu+kf^\mu-f_\nu f^\nu f^\mu=0, \label{cf3dIC}
\end{equation}
closely resembling the gravitational kink equation (\ref{gk2dEq(b)}). The
similarity between the sets of equations (\ref{CurvatureConstraint}),
(\ref{cf3dTE}), (\ref{cf3dIC}) and (\ref{gk2dEq(a)}), (\ref{gk2dEq(c)}),
(\ref{gk2dEq(b)}) is made even stronger by properly adapted coordinates.

\section{Darboux coordinates}
 \label{Dc}
By an appropriate coordinate transformation we now set $f^0=f^1=0$, i.e.
$F_{12}=F_{20}=0$. The existence of such a coordinate frame is
guarantied by a classical result in symplectic geometry, Darboux theorem
\cite{Darboux}. In 3d it ensures the possibility of finding, in a
\emph{finite} neighborhood of every point, local coordinates in such a way
that a given closed two-form (a $U(1)$ gauge field) can be rewritten, e.g.,
as
\begin{equation}
F_{\mu\nu}= \left(\begin{array}{cc}
\textrm{F}_{ij}\,&\,0\\
0\,&\,0
\end{array}\right)
\end{equation}
with $i,j=0,1$ and $\textrm{F}_{ij}$ a closed 2d two-form. Darboux
coordinates are determined up to the reparametrization $x^i\rightarrow
x'^i(x^0,x^1)$, $x^2\rightarrow x'^2(x^0,x^1,x^2)$, so that we still have a
certain freedom in choosing them. In the adapted frames the metric takes an
arbitrary form. Without loss of generality it can be parameterized as
\begin{equation}
g_{\mu\nu} dx^\mu dx^\nu = \textrm{g}_{ij} dx^i dx^j
+\textrm{h}\,(\textrm{a}_i dx^i + dx^2)^2, \label{MetricDc}
\end{equation}
where $\textrm{g}_{ij}$, $\textrm{a}_i$ and $\textrm{h}$, are arbitrary functions
of all coordinates.\footnote{We remark that, in spite of the similarity,
this is not a further Kaluza-Klein
Ansatz as the metric entries depend on the `extra' coordinate $x^2$.} Under
the residual covariance group, $\textrm{g}_{ij}$ transforms like a 2d metric
tensor, $\textrm{a}_i$ identifies with a 2d gauge potential taking values in
the 1d diffeomorphism algebra, while $\textrm{h}$ transforms like a scalar.
For $f^\mu$ space-like, $\textrm{g}_{ij}$ is Minkowskian, $\textrm{h}$
definite positive and time can be identified with $x^0$. For $f^\mu$
time-like, $\textrm{g}_{ij}$ is Euclidean, $\textrm{h}$ definite negative
and time has to be identified with $x^2$. In any case, we proceed as in
\S\ref{Intro} and re-express the effectively 2d gauge field
$\textrm{F}_{ij}=\partial_i\textrm{A}_j-\partial_j\textrm{A}_i$ in terms of
its dual scalar\footnote{With the only exceptions of coordinates and of the
gauge field $f$, 2d quantities are expressed in Roman characters.}
\begin{equation}
 \textrm{F}_{ij}\equiv \sqrt{|\textrm{g}|}\,\varepsilon_{ij}\,f.
\end{equation}
Consequently, in adapted coordinates, the 3d gauge field $f^\mu$ rewrites as
\begin{equation}
 f^\mu=\left(0,0,|\textrm{h}|^{-\frac{1}{2}}f\right).
\end{equation}
We also introduce the gauge curvature $ \textrm{f}_{ij}$ associated to the,
in general, non-abelian vector potential $\textrm{a}_i$
\begin{equation}
 \textrm{f}_{ij}= \partial_i\textrm{a}_j-\partial_j\textrm{a}_i
 -\textrm{a}_i\partial_2\textrm{a}_j+\textrm{a}_j\partial_2\textrm{a}_i
\end{equation}
and the relative dual scalar
\begin{equation}
 \textrm{f}_{ij}\equiv \sqrt{|\textrm{g}|}\,\varepsilon_{ij}\,\textrm{f}.
\end{equation}
Eventually, we take advantage of residual covariance by rescaling $x^2$
in such a way that
\begin{equation}
\textrm{h}=\textrm{h}(x^0,x^1).
\end{equation}
Correspondingly, covariance is reduced to arbitrary redefinitions of $x^i$,
$x^i\rightarrow x'^i(x^0,x^1)$, and linear redefinitions of $x^2$,
$x^2\rightarrow \xi(x^0,x^1)x^2$.

We now proceed by rewriting equations (\ref{cf3dEq}),
(\ref{cf3dTE}) and (\ref{cf3dIC}) in adapted Darboux coordinates.
The most convenient starting point is  equation (\ref{cf3dEq(b)}).
Its $ij$, $i2$ and $22$ components can be rearranged into the
lower dimensional covariant equations
\begin{subequations}
\begin{eqnarray}
&&\partial_2\textrm{g}_{ij}=0,\label{cf3dEq(b)Dc(a)}\\
&&\partial_2\textrm{a}_i= \mbox{$\frac{1}{2}$} \partial_i\ln(|\textrm{h}|
f^{-2}),
\label{cf3dEq(b)Dc(b)}\\
&&\partial_2 f=0.
\label{cf3dEq(b)Dc(c)}
\end{eqnarray}
\label{cf3dEq(b)Dc}
\end{subequations}
The first and third clearly require
\begin{equation}
\textrm{g}_{ij}=\textrm{g}_{ij}(x^0,x^1),\hskip0.3cm
f=f(x^0,x^1),\label{gij,f}
\end{equation}
while the integration of the second one produces
\begin{equation}
\textrm{a}_i=\mbox{$\frac{1}{2}$}
\partial_i\ln(|\textrm{h}|f^{-2})\,x^2+\bar{\textrm{a}}_i(x^0,x^1)
\nonumber
\end{equation}
with $\bar{\textrm{a}}_i(x^0,x^1)$ an arbitrary 2d one-form. The gauge
potential, $\textrm{a}_i$, is at most linear in the `extra' coordinate
$x^2$. As a consequence, we can take further advantage of residual
covariance to eliminate the $x^2$ dependence. In fact, the coordinate
transformation $x^2\rightarrow
\frac{1}{2}\ln(|\textrm{h}|f^{-2})\,x^2$ causes the off-diagonal
blocks of the metric to transform as $\textrm{a}_i\rightarrow
\textrm{a}_i-\frac{1}{2}\partial_i\ln(|\textrm{h}|f^{-2})\,x^2$,
making the transformed gauge potential to be independent of $x^2$,
\begin{equation}
\textrm{a}_i=\textrm{a}_i(x^0,x^1).\label{ai}
\end{equation}
In the new adapted coordinates we have $\partial_2 \textrm{a}_i=0$ so
equation (\ref{cf3dEq(b)Dc(b)}) implies the proportionality between
$\textrm{h}$ and $f^2$. The relative scale factor can be fixed to any
non-zero value by a constant rescaling of $x^2$. It is positive/negative
when $f^\mu$ is space/time-like. For definiteness, we fix the scale factor to be equal to $\pm 2$, i.e.
\begin{equation}
\textrm{h}=\,\pm 2\,f^2.\label{lambda}
\end{equation}
Here and in the following, upper/lower signs refer to space/time-like $f^\mu$.
The `extra' coordinate $x^2$ is now completely fixed and residual
covariance is reduced to 2d general covariance, $x^i\rightarrow
x'^i(x^0,x^1)$.\\
Next, we consider the integrability condition (\ref{cf3dIC}). Taking
(\ref{gij,f}), (\ref{ai}) and (\ref{lambda}) into account, its $i$ and $2$
components become
\begin{subequations}
\begin{eqnarray}
&&\partial_i\,(f^3\,\textrm{f})=0,\label{cf3dICDc(a)}\\
&&\textrm{D}^2f+k\,f\mp f^3+f^3\,\textrm{f}^2=0,
\label{cf3dICDc(b)}
\end{eqnarray}\label{cf3dICDc}\end{subequations}
with $\textrm{D}^2=\textrm{D}_i\textrm{D}^i$ and $\textrm{D}_i$
the covariant derivative associated to the 2d metric
$\textrm{g}_{ij}$. The first equation requires the product
$f^3\,\textrm{f}$ to be constant, allowing to reexpress
$\textrm{f}$ in terms of $f$. We set
\begin{equation}
f^3\,\textrm{f}=l\label{cf3dICDc(a)'}
\end{equation}
with $l$ an arbitrary constant. Consequently, (\ref{cf3dICDc(b)}) reduces to
\begin{subequations}
\begin{equation}
\textrm{D}^2f+k\,f\mp f^3+\frac{l^2}{f^3}=0.\label{gk3dEq(b)}
\end{equation}
For space-like $f^\mu$, (\ref{gk3dEq(b)})
is the gravitational kink equation (\ref{gk2dEq(b)}) augmented by a centrifugal
term of `angular momentum' $l$.
Eventually, we consider equations (\ref{cf3dEq(a)}) and (\ref{cf3dTE}) in
view of (\ref{gij,f}), (\ref{ai}), (\ref{lambda}) and (\ref{cf3dICDc(a)'}).
Equation (\ref{cf3dEq(a)}) reduces to
\begin{equation}
\textrm{R}=k \mp 3\,f^2-\frac{3l^2}{f^4},\label{gk3dEq(a)}
\end{equation}
for $\textrm{R}$ the 2d scalar curvature associated to $\textrm{g}_{ij}$.
For space-like $f^\mu$, (\ref{gk3dEq(a)}) corresponds to the curvature
constraint (\ref{gk2dEq(b)}) up to the `centrifugal' term $-3l^2/f^{4}$.
Equation (\ref{cf3dTE}) reduces instead to
\begin{equation}
\textrm{D}_i\textrm{D}_jf-\mbox{$\frac{1}{2}$}\textrm{g}_{ij}\textrm{D}^2f=0
\label{gk3dEq(c)}
\end{equation}\label{gk3dEq}\end{subequations}
corresponding to the traceless equation (\ref{gk2dEq(c)}) without any
modification.\\ Equations (\ref{gk3dEq}) are completely equivalent to equations
(\ref{cf3dEq}).

\section{Centrifugal kinks}
 \label{gk3d}
We now specialize to space-like gauge fields. For $l=0$ equations
(\ref{gk3dEq}) exactly correspond to the gravitational kink of equations
(\ref{gk2dEq}). Therefore, they support a 3d gravitational kink, with
$\textrm{g}_{ij}$ given by (\ref{gk2dMetric}), $\textrm{A}_i$ by
(\ref{gk2dVectorPotential}), $\textrm{a}_i=0$ and $\textrm{h}$ equal to
twice the square of (\ref{gk2dKinkProfile}). The resulting 3d spacetime is a
warped product of the 2d kink spacetime with the real line, where the warp
factor is twice the square of the kink profile. For $l\neq0$, the
biquadratic potential (\ref{VKink}) is supplemented by the centrifugal term
$l^2/(2f^2)$. Correspondingly, the scalar curvature $\textrm{R}$ is
augmented by $-3l^2/f^4=-(l^2/(2f^2))''$ (see Appendix B of
Ref.~\cite{Guralnik&Iorio&Jackiw&Pi03}). By properly choosing the
integration constant the new potential can be written as
\begin{equation}
V(f)=\frac{(\kappa-\lambda-f^2)^2(f^2+\lambda)}{4f^2},
\label{VCentrifugalKink}
\end{equation}
with the constants $\kappa$ and $\lambda$ related to $k$ and $l$
by $k=\kappa-3\lambda/2$ and
$2l^2=\left(\kappa-\lambda\right)^2\lambda$. The integration of
the corresponding flat space equation (see Appendices A and B of
Ref.~\cite{Guralnik&Iorio&Jackiw&Pi03}) is immediate and leads to
the following deformation of the general covariant kink
\begin{subequations}
\begin{eqnarray}
 &&\textrm{g}_{ij} dx^i dx^j =
   \mbox{$-\frac{\kappa^3\,\textrm{sech}^4\left(\frac{\sqrt{\kappa}}{2}x^1\right)\,
    \tanh^2\left(\frac{\sqrt{\kappa}}{2}x^1\right)}{
    4\kappa\,\tanh^2\left(\frac{\sqrt{\kappa}}{2}x^1\right)-4\lambda}$}(dx^0)^2+(dx^1)^2,\label{gk3dSolution(a)}\\
 &&\textrm{a}_i dx^i=
    \mbox{$\pm\frac{\kappa\sqrt{2\lambda}}{4(\kappa-\lambda)
     \cosh^2\left(\frac{\sqrt{\kappa}}{2}x^1\right)-4\kappa}$}dx^0,\label{gk3dSolution(b)}\\
 &&\textrm{h}=\mbox{$2\kappa\,\tanh^2\left(\frac{\sqrt{\kappa}}{2}x^1\right)-2\lambda$},\label{gk3dSolution(c)}\\
 &&\textrm{A}_i dx^i=
    \mbox{$\pm\frac{\kappa}{2}\,{\textrm{sech}^2\left(\frac{\sqrt{\kappa}}{2}x^1\right)}$}dx^0,\label{gk3dSolution(d)}
\end{eqnarray}
with the corresponding centrifugal distortion of the kink profile
\begin{equation}
f=\mbox{$\pm\sqrt{\kappa\tanh^2\left(\frac{\sqrt{\kappa}}{2}x^1\right)-\lambda}$}\label{gk3dSolution(e)}.
\end{equation}\label{gk3dSolutions}\end{subequations}
For $l=0$ the constants $\kappa$ and $\lambda$ respectively reduce
to $k$ and $0$, so that (\ref{gk3dSolutions}) correctly reproduces
the 3d gravitational kink. For $l\neq0$ the 2d metric
$\textrm{g}_{ij}$, and the corresponding scalar curvature
$\textrm{R}$, is singular at
$x^1=\pm\frac{2}{\sqrt{\kappa}}\mbox{arctanh}\sqrt{\frac{\lambda}{\kappa}}$,
while the gauge field $f$ is only defined for
$|x^1|>\frac{2}{\sqrt{\kappa}}\mbox{arctanh}\sqrt{\frac{\lambda}{\kappa}}$.
The effect of the centrifugal interaction is that of breaking the
kink in two parts, pushing them apart, and thus dividing spacetime
in two disconnected regions.

\section*{Acknowledgments}

We would like to thank  Daniel Grumiller and Roman Jackiw for inspiring
conversations. This work was supported by the Royal Society.

\end{document}